\documentclass[12pt,preprint]{emulateapj}
\usepackage{multirow}
\usepackage[usenames,dvipsnames]{xcolor}

\shorttitle{\textsc{Empirical line lists and absorption cross sections for methane at high temperature}}
\shortauthors{\textsc{Hargreaves et al. 2015}}

\begin{document}

\title{Empirical line lists and absorption cross sections for methane at high temperature}


\author{R. J. Hargreaves$^{1}$, P. F. Bernath$^{1}$, J. Bailey$^{2}$, \& M. Dulick$^{1}$}
\affil{$^{1}$Department of Chemistry \& Biochemistry, Old Dominion University, 4541 Hampton Boulevard, Norfolk, VA 23529}
\affil{$^{2}$School of Physics, University of New South Wales, NSW 2052, Australia}
\email{\textrm{Email:} rhargrea@odu.edu}
\email{\textrm{Note: Online data is available from the Astrophysical Journal}}

\begin{abstract}
Hot methane is found in many ``cool" sub-stellar astronomical sources including brown dwarfs and exoplanets, as well as in combustion environments on Earth. We report on the first high-resolution laboratory absorption spectra of hot methane at temperatures up to 1200 K. Our observations are compared to the latest theoretical spectral predictions and recent brown dwarf spectra. The expectation that millions of weak absorption lines combine to form a continuum, not seen at room temperature, is confirmed. Our high-resolution transmittance spectra account for both the emission and absorption of methane at elevated temperatures. From these spectra, we obtain an empirical line list and continuum that is able to account for the absorption of methane in high temperature environments at both high and low resolution. Great advances have recently been made in the theoretical prediction of hot methane, and our experimental measurements highlight the progress made and the problems that still remain.
\end{abstract}




\section{INTRODUCTION}

Methane (CH$_{4}$) is the simplest hydrocarbon molecule and the main constituent of natural gas. It plays a central role in combustion and, along with carbon dioxide, is a major cause of climate change \citep{petit1999}. In addition to its terrestrial importance, absorption bands of CH$_{4}$ feature prominently in the infrared spectra of numerous astrophysical objects over a broad temperature range. The onset of CH$_{4}$ absorption appears in sub-stellar objects below temperatures of $\sim$2000 K \citep{kirkpatrick2005}. Indeed the spectrum of hot CH$_{4}$ is a defining feature for characterizing brown dwarfs, objects which do not fuse hydrogen in their core and have masses between those of stars and planets. The T-dwarfs are characterized by the strengthening of CH$_{4}$ absorption in the near-infrared \citep{burgasser2006}; thus these objects are sometimes referred to as ``methane'' dwarfs \citep{hauschildt2009}. CH$_{4}$ remains a prominent absorbing species in astrophysical spectra down to $\sim$80 K, the atmospheric temperatures of the Outer Planets and Titan \citep{atreya2003, flaser2005}. Furthermore, CH$_{4}$ has been detected in the atmosphere of exoplanets known as hot-Jupiters \citep{swain2008} and it is anticipated that CH$_{4}$ detection can be used as a key biosignature for terrestrial exoplanets \citep{desmarais2002}.

As a consequence of the ubiquity of CH$_{4}$, the infrared spectrum has been the focus of numerous spectroscopic and theoretical studies, primarily at room temperature \citep[e.g.,][]{albert2009, brown2013, campargue2013}. The HITRAN database \citep{rothman2013} compiles the most recent line parameters of CH$_{4}$ and is intended for the remote sensing of the Earth's atmosphere \citep{rinsland2009}, but it is often also used in astronomy \citep{maltagliati2015}.  However, the complexity of an infrared molecular spectrum (i.e., observed vibration-rotation lines) considerably increases with higher temperatures due to the population of excited levels. These high-temperature transitions are weak in cold spectra and are therefore not observed. For this reason, they are not usually included in resulting ``cold'' line lists, such as HITRAN. When these line lists are applied to high-temperature brown dwarf models, it comes as no surprise that CH$_{4}$ absorption is underestimated due to these missing high-temperature transitions \citep{bailey2012}. The use of high temperature data is not solely limited to astrophysical observations as CH$_{4}$ spectral features can also be used to monitor combustion processes over similar temperature regimes \citep{sur2015, jourdanneau2007}.

With the escalation of interest in brown dwarfs and hot-Jupiters arising from improved detection methods, it has been necessary to provide data of CH$_{4}$ at elevated temperatures for comparison \citep{nassar2003, perrin2007, thievin2008, hargreaves2013}. So far, the infrared spectrum of CH$_{4}$ has proved to be a difficult challenge for theory because of the high molecular symmetry and polyad structure arising from close-lying vibrational level interactions \citep{boudon2006}. Nevertheless, advances are being made in assigning experimental data and recently, two state-of-the-art theoretical CH$_{4}$ line lists have been published by \citet{rey2014} as well as by \citet{yurchenko2014a} as part of the ExoMol project \citep{tennyson2012}. Currently, these two line lists provide the best broadband comparisons with high temperature environments and are referred to herein as 10to10 \citep{yurchenko2014a} and RNT \citep{rey2014}. For example, the 10to10 CH$_{4}$ line list (so named because it contains approximately 10 billion lines) has been shown to provide good agreement with observed hot CH$_{4}$ emission during the Shoemaker-Levy 9 impact of Jupiter as well as with the CH$_{4}$ opacity of a brown dwarf \citep{yurchenko2014b}. In both cases these spectra have effective temperatures in excess of 1000 K and require hot line lists for analysis.

However, despite these recent advances, experimental observations provide the definitive description of CH$_{4}$ absorption at high temperatures and are therefore crucial to refining theoretical predictions and matching observation. {\color{Black}For astronomical purposes, it has been shown that empirical line lists produced over the appropriate temperatures can be used for molecular spectra in the submillimeter region \citep{fortman2010a}. This method is particularly useful for so-called astrophysical ``weeds'' \citep{fortman2010b} and methanol \citep{mcmillan2014} where traditional spectroscopic methods cannot keep pace with current observations.}

The temperature regime of brown dwarf and exoplanets can be achieved in the laboratory and until now, only high-temperature infrared emission spectra of CH$_{4}$ have been available at high resolution over the appropriate temperature range \citep{nassar2003, thievin2008, hargreaves2013}. This article reports the first such absorption spectra of CH$_{4}$ at high temperatures (up to 900$^{\circ}$C). Additionally, we demonstrate the necessary inclusion of a quasi-continuum that is required to replicate observation using our empirical line lists. The inclusion of our data into brown dwarf models compares well with observations and with models that use theoretical line lists containing millions, or even, billions of lines.

\section{EXPERIMENTAL MEASUREMENTS}

Nine CH$_{4}$ transmittance spectra have been acquired between room temperature (i.e., 23$^{\circ}$C) and 200 $–-$ 900$^{\circ}$C in 100$^{\circ}$C intervals. For each transmittance spectrum, four individual spectra are required to correct for the emission of CH$_{4}$ and the cell. At each temperature $T$, the four spectra are as follows:
\begin{enumerate}
  \item $A_{\scriptsize{\textrm{ab}}}$	-- CH$_{4}$ absorption at $T$ against the background of $A_{\scriptsize{\textrm{ref}}}$.
  \item $A_{\scriptsize{\textrm{ref}}}$	-- Background broadband emission from the external source (including $B_{\scriptsize{\textrm{ref}}}$).
  \item $B_{\scriptsize{\textrm{em}}}$	-- CH$_{4}$ emission at $T$ against the background of $B_{\scriptsize{\textrm{ref}}}$.
  \item $B_{\scriptsize{\textrm{ref}}}$	-- Background thermal emission of the cell at $T$.
\end{enumerate}

\begin{figure*}[]
\centering
\includegraphics[scale=.65]{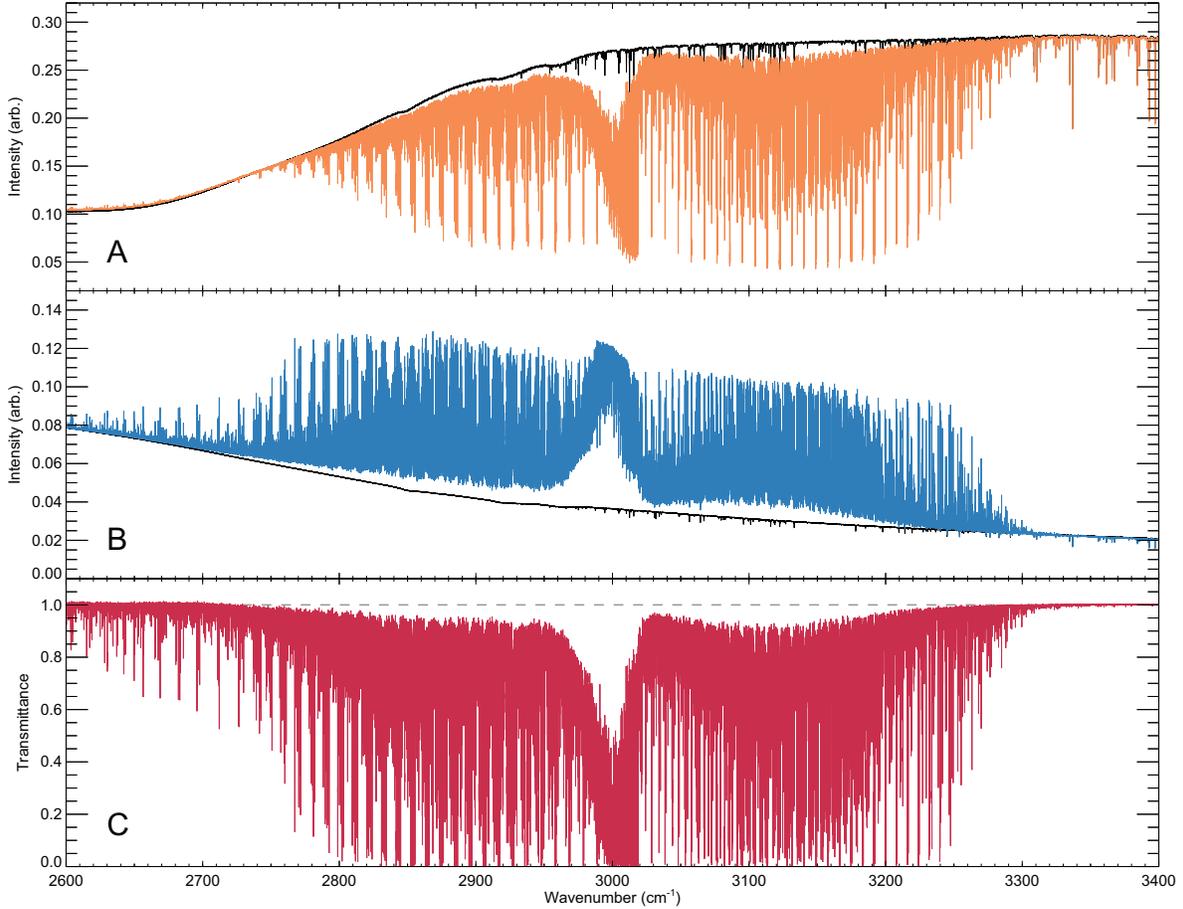}
\caption{Emission, absorption and transmittance spectra of CH$_{4}$ at 700$^{\circ}$C for the 3 $\mu$m region. (A) CH$_{4}$ absorption from external lamp ($A_{\scriptsize{\textrm{ab}}}$, orange) and background due to external lamp emission ($A_{\scriptsize{\textrm{ref}}}$, black). (B) CH$_{4}$ emission ($B_{\scriptsize{\textrm{em}}}$, blue) with a baseline due to thermal emission of the cell ($B_{\scriptsize{\textrm{ref}}}$, black). (C) Transmittance of CH$_{4}$ at 700$^{\circ}$C after combining both emission and absorption from the cell as $\tau = (A_{\scriptsize{\textrm{ab}}} -– B_{\scriptsize{\textrm{em}}}) / (A_{\scriptsize{\textrm{ref}}} –- B_{\scriptsize{\textrm{ref}}}$). The CH$_{4}$ quasi-continuum is illustrated by the deviation of the transmittance baseline from unity (dashed), most obvious in the Q-branch near 3000 cm$^{-1}$ (3.3 $\mu$m).\label{fig1} \vspace{0.5cm}}
\end{figure*}

A total of 36 spectra were acquired and the four resultant spectra at 700$^{\circ}$C are illustrated in Figure~\ref{fig1}. The 2600 $–-$ 5000 cm$^{-1}$ (2.0 $-–$ 3.8 $\mu$m) spectral range has been covered and includes pentad and octad regions of the CH$_{4}$ infrared spectrum. Experimental parameters for all measurements are provided in Table~\ref{tab1}.

\begin{deluxetable}{lc}
\tabletypesize{\small}
\tablecolumns{2}
\tablewidth{0pt}
\tablecaption{Experimental conditions and Fourier transform parameters \label{tab1}}
\tablehead{\colhead{Parameter} & \colhead{Value\tablenotemark{a}} }
\startdata
Temperature range                     & 23 $-$ 900$^{\circ}$C \\
CH$_{4}$ pressure                     & 60 Torr\tablenotemark{b} \\
Scans                                 & 600 \\
Sample cell material                  & Quartz (SiO$_{2}$) \\
External source                       & Tungsten halogen lamp\tablenotemark{c} \\
\multirow{2}{*}{Detector}             & Indium antimonide (InSb) \\
                                      & (response $>$ 1800 cm$^{-1}$) \\
Beam splitter                         & Calcium Fluoride (CaF$_{2}$) \\
\multirow{2}{*}{Spectrometer windows} & CaF$_{2}$ \\
                                      & (transmission $>$ 1300 cm$^{-1}$) \\
\multirow{2}{*}{Filter}               & Germanium  \\
                                      & (transmission $<$ 5600 cm$^{-1}$) \\
Resolution                            & 0.02 cm$^{-1}$ \\
Aperture                              & 2.0 mm \\
Apodization function                  & Norton-Beer weak \\
Phase correction                      & Mertz \\
Zero-fill factor                      & $\times$16 \\
\enddata
\tablenotetext{a}{\footnotesize{all spectra recorded under same conditions except where stated.}}
\tablenotetext{b}{\footnotesize{for $A_{\textrm{\tiny{ab}}}$ and $B_{\textrm{\tiny{em}}}$.}}
\tablenotetext{c}{\footnotesize{no external source for $B_{\textrm{\tiny{em}}}$ and $B_{\textrm{\tiny{ref}}}$. \vspace{0.5cm}}
}
\end{deluxetable}

\begin{figure}[b!]
\centering
\includegraphics[scale=.60]{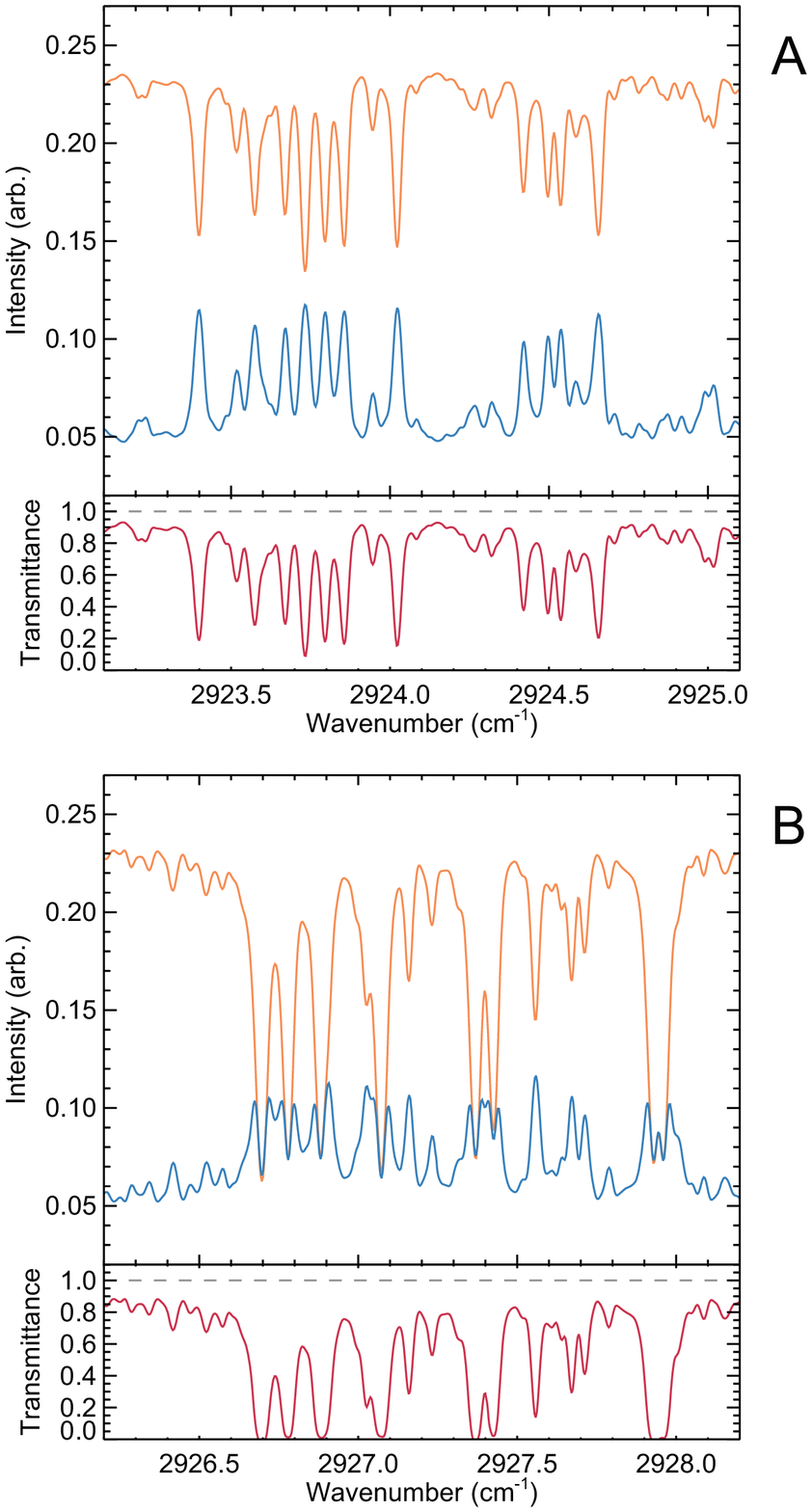}
\caption{Emission correction for optically thin (A) and optically thick (B) lines of CH$_{4}$ at 700$^{\circ}$C. The upper panels display the absorption (orange) and emission (blue) spectrum on the same arbitrary intensity scale; the lower panels display the transmittance (red).\label{fig2}}
\end{figure}

A 0.5 m sample cell was heated by a tube furnace and the transmittance spectra were recorded using an updated Bruker IFS 120 HR Fourier transform infrared spectrometer. The sample cell is made entirely of quartz (including quartz windows) and is contained completely within the heated portion of the furnace. This avoids the problem of absorption effects seen in previous emission studies  \citep{hargreaves2013, hargreaves2011, hargreaves2012} caused by cold CH$_{4}$ found in the ends of the cell beyond the heated region. The cell was aligned with the entrance aperture of the spectrometer and the radiation was focused into the instrument using a calcium fluoride lens. CH$_{4}$ absorption spectra and backgrounds ($A_{\scriptsize{\textrm{ab}}}$ and $A_{\scriptsize{\textrm{ref}}}$) were obtained by using a tungsten halogen lamp as an external emission source (broad spectral coverage above $\sim$2600 cm$^{-1}$). CH$_{4}$ emission spectra and backgrounds ($B_{\scriptsize{\textrm{em}}}$ and $B_{\scriptsize{\textrm{ref}}}$) were observed by switching the lamp off. To produce an emission-corrected transmittance spectrum $\tau$, the spectra are combined as
\begin{equation}
\label{eqn1}
\tau = \frac{A_{\scriptsize{\textrm{ab}}} -– B_{\scriptsize{\textrm{em}}}}{A_{\scriptsize{\textrm{ref}}} –- B_{\scriptsize{\textrm{ref}}}},
\end{equation}
as shown in Figure~\ref{fig1}. At high temperature, the emission component of CH$_{4}$ can be of comparable strength to the absorption component when experimental conditions remain unchanged. Figure~\ref{fig2} displays the effect of the emission correction for optically thin and optically thick lines of CH$_{4}$. For optically thin lines, the resulting emission-corrected transmittance is a true measure of the absorption depth. This is more clearly demonstrated when considering optically thick lines. In this case, the lines are saturated, and now the emission correction results in a transmittance of zero (i.e., $\tau = 0$). By comparing both the emission and absorption on the same scale, these lines are seen to be saturated (i.e., equal). The transmittance is then correctly calculated to be zero as expected from Kirchhoff's law of thermal radiation that states emissivity and absorptivity are equal. If the absorption or emission spectra are observed independently, then each spectrum will appear distorted, as previously observed \citep{hargreaves2013}.

The CH$_{4}$ molecules begin to decompose near 1000$^{\circ}$C and as a consequence, deposits build up within the quartz cell. These deposits caused clouding of the cell windows and significantly reduced the transmission at temperatures greater than 900$^{\circ}$C; therefore reliable measurements at 1000$^{\circ}$C were not achieved.

\section{EMPIRICAL CH$_{4}$ LINE LISTS}

WSpectra \citep{carleer2001} was used to measure line positions and intensities of all absorption lines. These lists were wavenumber and intensity calibrated with strong lines found in the HITRAN 2012 database \citep{rothman2013} and line positions are accurate to within $\pm$0.002 cm$^{-1}$. In small sections of the transmission spectra, water line residuals have been removed when cancellation was not complete.

The CH$_{4}$ lower state energies are calculated by comparing the calibrated intensities of lines at each temperature \citep{hargreaves2013, hargreaves2011, hargreaves2012}. Line intensities can be calculated by using
\begin{equation}
\label{eqn2}
S' = \frac{2\pi^{2}\nu S_{J'J''}}{3\varepsilon_{0}hcQ} \exp \left(⁡-\frac{E''}{kT}\right) \left[ 1-\exp\left(⁡-\frac{h\nu}{kT}\right) \right]
\end{equation}
where $\nu$ is the line frequency, $S_{J'J''}$ is the line strength, $\varepsilon_{0}$ is the permittivity of free space, $h$ is the Planck constant, $c$ is the speed of light, $Q$ is the internal partition function, $E''$ is the lower state energy, $k$ is the Boltzmann constant and $T$ is the temperature \citep{bernath2005}. However, by comparing the intensity of the same line at two temperatures, the equation simplifies to
\begin{equation}
\label{eqn3}
\frac{S'}{S'_{0}} = \frac{Q_{0}}{Q}  \exp\left(\frac{E''}{kT_{0}}⁡-\frac{E''}{kT}\right) \left[\frac{1-\exp⁡(-\frac{h\nu}{kT})}{1-\exp⁡(-\frac{h\nu}{kT_{0}})}\right],
\end{equation}
where the zero subscript refers to the reference temperature at which all intensities are compared. For these results, the lowest temperature at which the line appears is used as the reference. The CH$_{4}$ partition function for this work is taken from \citet{laraia2011}.

Equation~\ref{eqn3} can be rearranged as
\begin{equation}
\label{eqn4}
\ln⁡ \left( \frac{S'QR_{0}}{S'_{0}Q_{0}R}\right) = \frac{E''}{kT_{0}} - \frac{E''}{kT},
\end{equation}
where $R = 1 - \exp(- h\nu / kT)$ and $R_{0} = 1 - \exp⁡(-h\nu / kT_{0})$. Plotting $\ln(S'QR_{0}/S'_{0}Q_{0}R)$ against $1/kT$ yields the lower state energy $E''$ as the slope.

A small proportion of spectral lines are completely saturated in all of our transmittance spectra, therefore the intensity and positions cannot be obtained. In order to have a complete line list that can be used to match observation, it is necessary to add line parameters to our line lists from HITRAN. Care was taken to avoid adding unnecessary HITRAN lines back into our empirical line lists by using appropriate intensity limits. These relate to a maximum intensity above which saturation effects are observed. The HITRAN lines have been calculated at the appropriate temperature using Equation~\ref{eqn3}, and only included if they are greater than $5.0\times10^{-22}$ cm molecule$^{-1}$ in the pentad region and $3.0\times10^{-21}$ cm molecule$^{-1}$ in the octad region. Due to these intensity limits, the vast majority of HITRAN additions are in pentad region. Table~\ref{tab2} provides the total number of lines at each temperature, the number of lines that have been added from HITRAN and the corresponding intensity sums.

\begin{deluxetable*}{cccccc}
\tabletypesize{\small}
\tablecolumns{6}
\tablewidth{0pt}
\tablecaption{Total lines and intensity sums for each temperature\label{tab2}}
\tablehead{ \colhead{Temperature}             & \colhead{Total} & \colhead{Intensity}            & \colhead{Number}                  & \colhead{Intensity sum}               & \colhead{Integrated} \\
\colhead{($^{\circ}$C)}  & \colhead{Lines} & \colhead{sum of}               & \colhead{of total}                & \colhead{of HITRAN}                  & \colhead{absorption} \\
                         &                 & \colhead{total lines}          & \colhead{lines added}             & \colhead{additions\tablenotemark{b}} & \colhead{cross section of} \\
                         &                 & \colhead{(cm molecule$^{-1}$)} & \colhead{from}                    & \colhead{(cm molecule$^{-1}$)}       & \colhead{observed continua}\\
                         &                 &                                & \colhead{HITRAN\tablenotemark{a}} &                                      & \colhead{(cm molecule$^{-1}$)}}
\startdata
200 & 19,329 & 1.19E-17 & 1268 & 1.05E-17 & 3.67E-19\\
300 & 22,391 & 1.15E-17 & 1733 & 1.01E-17 & 4.89E-19\\
400 & 24,046 & 1.10E-17 & 2047 & 9.46E-18 & 7.72E-19\\
500 & 25,508 & 1.04E-17 & 2295 & 8.68E-18 & 1.21E-18\\
600 & 25,904 & 9.55E-18 & 2391 & 7.73E-18 & 1.67E-18\\
700 & 26,388 & 8.59E-18 & 2370 & 6.69E-18 & 2.29E-18\\
800 & 26,199 & 7.60E-18 & 2309 & 5.64E-18 & 3.03E-18\\
900 & 24,345 & 6.67E-18 & 2218 & 4.65E-18 & 3.89E-18
\enddata
\tablenotetext{a}{\footnotesize{added from HITRAN 2012 \citep{rothman2013} and included in total line count.}}
\tablenotetext{b}{\footnotesize{only accounts for the HITRAN additions, not the total HITRAN intensity sum.} \vspace{0.5cm}}
\end{deluxetable*}

A quality code (QC) is provided for each empirical lower state energy $E''$. Each QC is determined from
\begin{equation}
\label{eqn5}
F =\frac{nR^{2}}{9}
\end{equation}
where $n$ refers to the number of temperature spectra that the line appears ($n_{\textrm{\scriptsize{max}}}$ = 9) and $R^{2}$ is the coefficient of determination of the empirical $E''$ calculation. QC = 1 indicates $F \geq 0.8$, QC = 2 is when $0.8 > F \geq 0.5$ and QC = 3 means $0.5 > F > 0.0$. When an $E''$ could not be calculated, $E''$ and QC are left blank. Line parameters that have been inserted from HITRAN are indicated by QC = H. The QC is an indication of the error of $E''$ and is strongly dependent upon $n$. For $n_{\textrm{\scriptsize{max}}}$, approximate accuracies of $E''$ between 0-20\%, 20-100\% and $>$100\% are estimated for QC = 1, 2 and 3, respectively. It should be noted that in cases where $n$ is low, these errors can be larger.

The empirical CH$_{4}$ line lists include temperature $T$ ($^{\circ}$C), line position $\tilde\nu$ (cm$^{-1}$), line intensity $S'$ (cm molecule$^{-1}$), lower state energy $E''$ (cm$^{-1}$) and quality code QC. A sample is provided in Table~\ref{tab3}, and the complete table is available online from the Astrophysical Journal.

\begin{deluxetable}{ccccc}[B!]
\tabletypesize{\small}
\tablecolumns{5}
\tablewidth{0pt}
\tablecaption{Line list format\tablenotemark{a} \label{tab3}}
\tablehead{\colhead{$T$} & \colhead{$\tilde\nu$}  & \colhead{$S'$}  & \colhead{$E''$}  &  \colhead{QC}\\
 \colhead{($^{\circ}$C)} & \colhead{(cm$^{-1}$)}  & \colhead{(cm molecule$^{-1}$)} & \colhead{(cm$^{-1}$)} &  }
\startdata
700 & 2979.716448 & 6.727E-22 & 1.367734E+03 & H \\
700 & 2979.754847 & 6.514E-22 & 1.368277E+03 & H \\
700 & 2979.833000 & 1.038E-22 & 2.240000E+03 & 1 \\
700 & 2979.891500 & 4.748E-23 & 1.586000E+03 & 2 \\
700 & 2979.913000 & 1.501E-23 & 1.234000E+04 & 3 \\
700 & 2979.937400 & 5.109E-23 & 5.214000E+02 & 1 \\
700 & 2980.012800 & 1.194E-21 & 1.369182E+03 & H \\
700 & 2980.070000 & 3.326E-22 & 1.848000E+03 & 1 \\
700 & 2980.124500 & 6.278E-23 & 4.502000E+03 & 2 \\
700 & 2980.161200 & 2.621E-22 & 2.239000E+03 & 1 \\
700 & 2980.204800 & 2.435E-22 & 3.223000E+03 & 2 \\
700 & 2980.277900 & 1.979E-22 & 1.745000E+03 & 1 \\
700 & 2980.312700 & 2.237E-23 &              &   \\
700 & 2980.340400 & 3.530E-23 & 3.786000E+03 & 3 \\
700 & 2980.376500 & 9.238E-23 & 4.017000E+03 & 3 
\enddata
\tablenotetext{a}{\footnotesize{full table is available online from the Astrophysical Journal.} \vspace{0.5cm}}
\end{deluxetable}

\section{QUASI-CONTINUUM CROSS SECTIONS}

The empirical line lists alone do not account for the complete absorption of CH$_{4}$ at high temperature. For particular regions of the transmittance spectrum, it is not possible to assign every feature because, in many instances, transitions are blended together. The most noticeable regions for this effect to occur are in Q-branch features, such as near 3000 cm$^{-1}$ in the $\nu_{3}$ band of CH$_{4}$ (Figure~\ref{fig1}). This is not an important issue for room temperature spectra and below, since the majority of lines can be observed in cold spectra or easily calculated. But at elevated temperatures, transitions from hot bands are not well known. The blending of these ``hot'' lines becomes very severe and an unresolvable quasi-continuum emerges. Many observed ``lines'' are actually blended features and have effective parameters in our line lists.

These continua are obtained from the line intensity output of the peak finding program WSpectra. Each line intensity is determined from a reference baseline and for perfect transmission spectra, this baseline is unity. However, WSpectra is able to calculate the line intensity as it appears from the observed continuum. This baseline was smoothed using the OPUS 7.0 spectral software from Bruker. The infrared absorption continua can then be calibrated using
\begin{equation}
\label{eqn6}
\sigma = - \xi\frac{10^{4}kT}{Pl} \textrm{ln}⁡(\tau)
\end{equation}
where $\sigma$ is the absorption cross section (cm$^{2}$ molecule$^{-1}$), $k$ is the Boltzmann constant, $T$ is the temperature (in K), $P$ is the pressure of the gas (in Pa), $l$ is the pathlength of the absorbing gas (in m), $\tau$ is the transmittance of the continuum and $\xi$ is a correction coefficient that is assumed to be unity for this study \citep{harrison2010}. For our measurements, the pressure was 8 kPa (i.e., 60 Torr) and the pathlength of the absorbing gas is taken as 0.5 m, the length of the cell.

\begin{figure*}[t!]
\centering
\includegraphics[scale=.8]{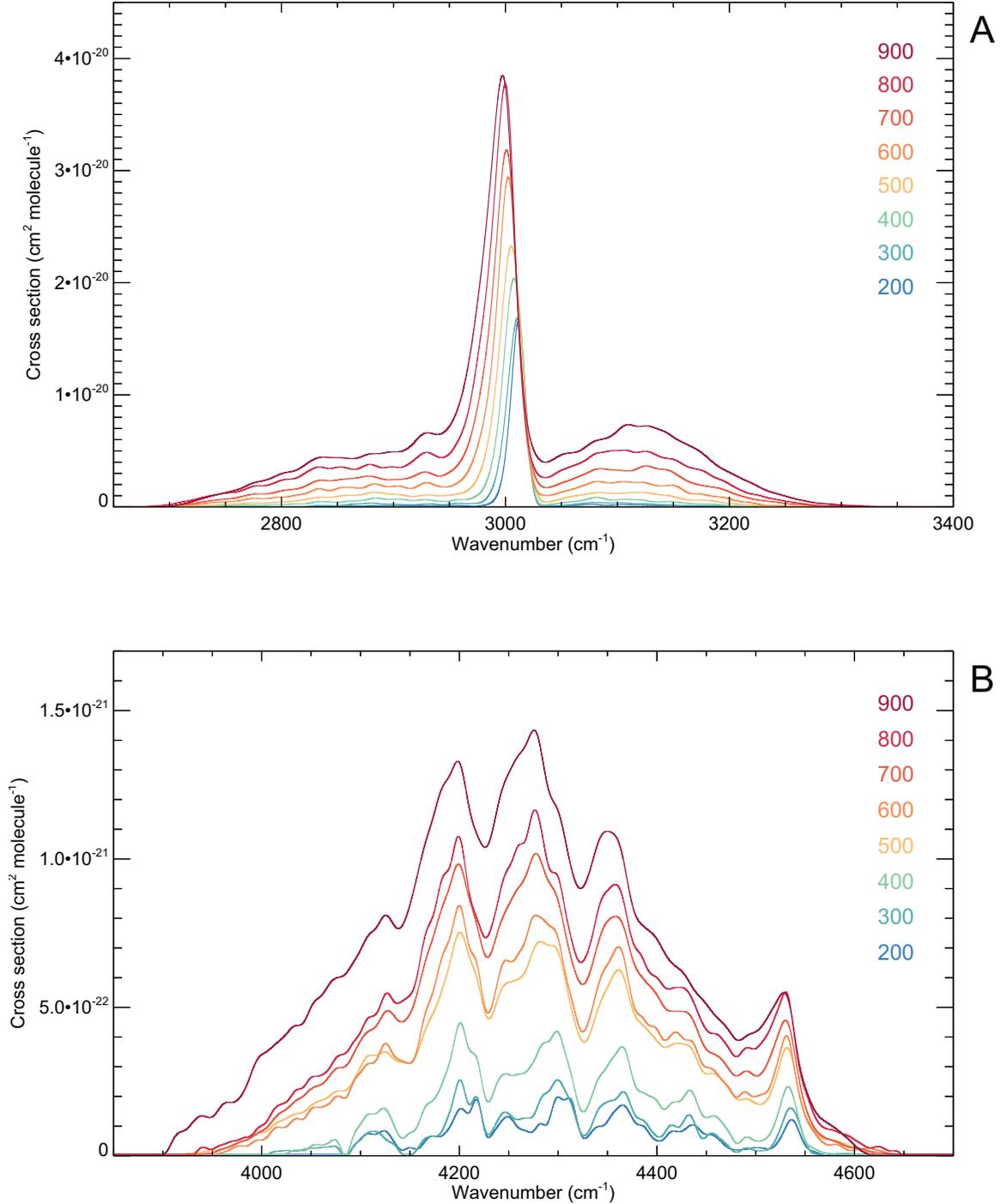}
\caption{The CH$_{4}$ continuum cross sections at increasing temperature for the pentad (A) and octad (B) regions.\label{fig3} \vspace{0.2cm}}
\end{figure*}

The quasi-continuum cross sections are displayed in Figure~\ref{fig3} and their integrated absorptions are provided in Table~\ref{tab2}. When the integrated absorptions are added to the total line intensity sums (also given in Table~\ref{tab2}), the full empirical contribution of CH$_{4}$ is given. For example, at 900$^{\circ}$C the empirical intensity sum (total lines plus continuum) is $1.06\times10^{-17}$ cm molecule$^{-1}$ compared with $5.73\times10^{-18}$ cm molecule$^{-1}$ for the total HITRAN intensity sum at 900$^{\circ}$C over the same spectral region. This demonstrates that the quasi-continuum cross sections account for many of the ``hot'' transitions that are not contained in HITRAN, and therefore must be used together with the empirical line lists. The empirical cross sections are provided in the standard HITRAN cross section format \citep{rothman2013} and are available online from the Astrophysical Journal, along with the CH$_{4}$ line lists. The presence of this continuum is predicted by the latest theoretical line lists but only appears if millions (or possibly billions) of individual lines are included. This work presents the first experimental observation of this continuum effect for CH$_{4}$.

\section{ANALYSIS OF CH$_{4}$ AT HIGH RESOLUTION}

\begin{figure*}[b!]
\centering
\includegraphics[scale=.8]{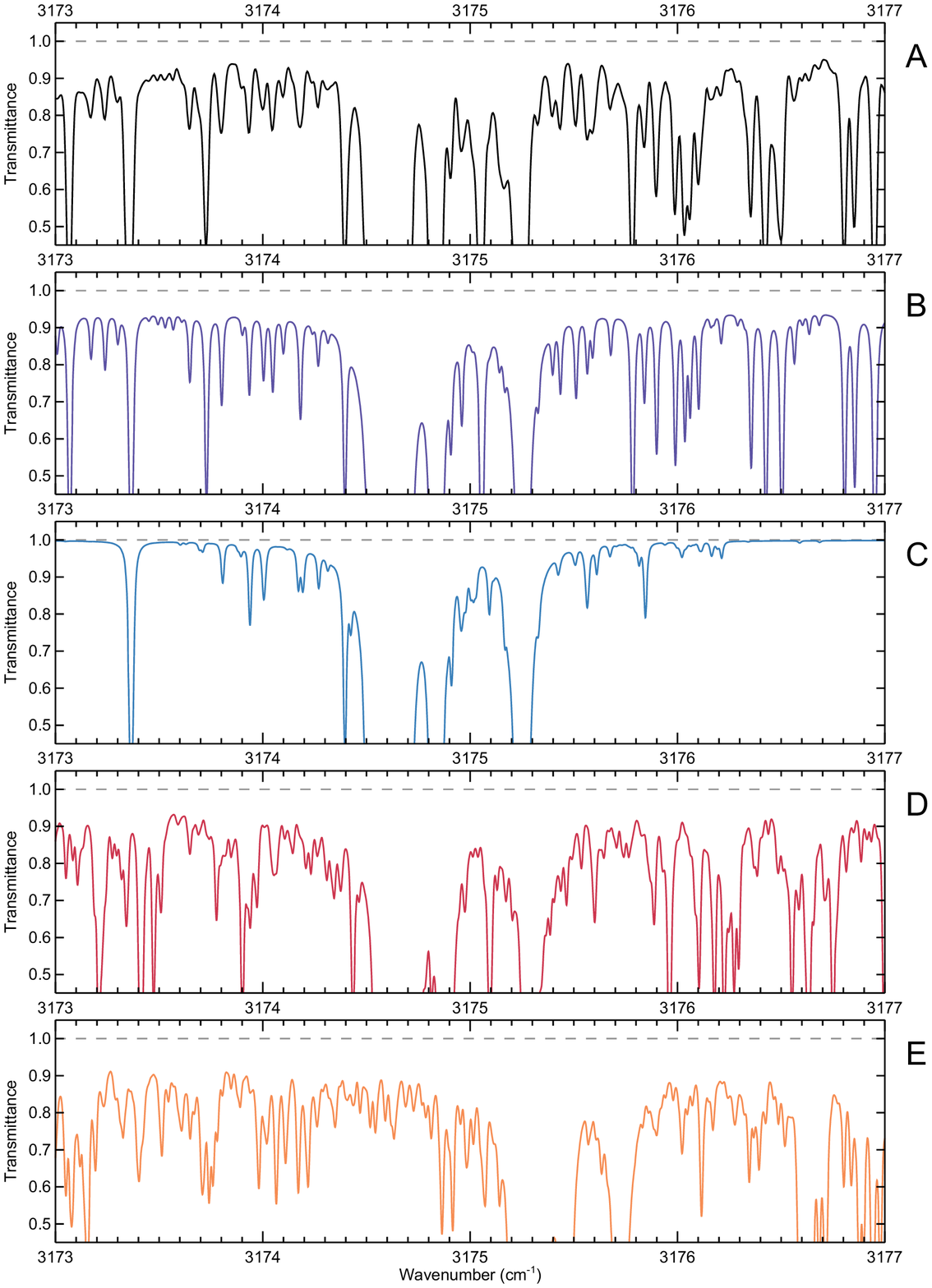}
\caption{High resolution comparison of laboratory measurements at 700$^{\circ}$C with calculated transmittance spectra using empirical and theoretical line lists. (A) The observed transmittance spectrum of CH$_{4}$. (B) Calculated transmittance spectrum using the new empirical CH$_{4}$ line list and CH$_{4}$ continuum cross section.  (C) Calculated transmittance spectrum using the HITRAN  CH$_{4}$ line list. (D) Calculated transmittance spectrum using the RNT CH$_{4}$ line list. (E) Calculated transmittance spectrum using the 10to10 CH$_{4}$ line list.\label{fig4}}
\end{figure*}

Figure~\ref{fig4} displays the observed CH$_{4}$ spectrum at 700$^{\circ}$C along with transmittance spectra calculated using theoretical and experimental CH$_{4}$ line lists. The transmittance spectra were calculated using RFM\footnote{Reference Forward Model, RFM (v4.30), A. Dudhia, University of Oxford \url{http://www.atm.ox.ac.uk/RFM} }, a line-by-line radiative transfer forward model. RFM is more commonly used for simulating atmospheric spectra but is able to calculate transmittance spectra through a specified pathlength. For comparison purposes, a pressure-pathlength containing CH$_{4}$ at a pressure of 60 Torr and length 0.15 m was used to obtain approximate absorption depths (resolution of $\sim$0.015 cm$^{-1}$). Average wavenumber dependent broadening parameters were obtained from the lines contained in HITRAN and are estimated to have a maximum error of 50\% for the region displayed in Figure~\ref{fig4}. These average broadening parameters were applied to the experimental line lists from this study, as well as the HITRAN, 10to10 and RNT CH$_{4}$ line lists. No instrument lineshape was included for the RFM simulations and the line lists were input into RFM at the elevated temperatures (as opposed to the usual temperature of 296 K). In Figure~\ref{fig4}, the 10to10 and RNT simulated spectra are at 1000 K whereas the observed spectrum, empirical line list and HITRAN are at 973 K (700$^{\circ}$C). The RNT CH$_{4}$ line list is provided at 1000 K and, to keep theoretical comparisons consistent, the 10to10 line list was also calculated at 1000 K. The difference between these temperatures is small and would only lead to slight changes in intensity for the lines shown. Although the RFM pathlength is not a true representation of experiment, a consistent treatment between all line lists allows high-resolution comparisons to be made.

Figure~\ref{fig4} demonstrates the presence of an underlying continuum since the observed spectrum does not reach unity. As expected, missing hot lines in the HITRAN database mean that it is incomplete at high temperature even if a continuum cross section were to be included. The theoretical line lists are able to account for the continuum effect at high temperatures and RNT produces a better match to experiment for the region shown. However, it is the line lists and continua presented in this article that show the best agreement with high resolution experimental spectra.

A maximum number of 26,388 lines are contained in the 700$^{\circ}$C line list and these lines cover the 2600 $-–$ 5000 cm$^{-1}$ region. These lines only account for a very small fraction of the total transitions that have been calculated to occur in this region. However, the majority of these missing lines will always remain below observational limits at elevated temperatures. The blending together of these weak missing lines, each broadened by the Doppler effect, leads to the high temperature absorption continuum. Although millions (and even billions) of lines can be calculated from theory, a continuum cross section and relatively small line list are able to reproduce the high resolution structure of CH$_{4}$.

\section{ANALYSIS OF CH$_{4}$ AT LOW RESOLUTION}

\begin{figure}[t!]
\centering
\includegraphics[scale=0.40]{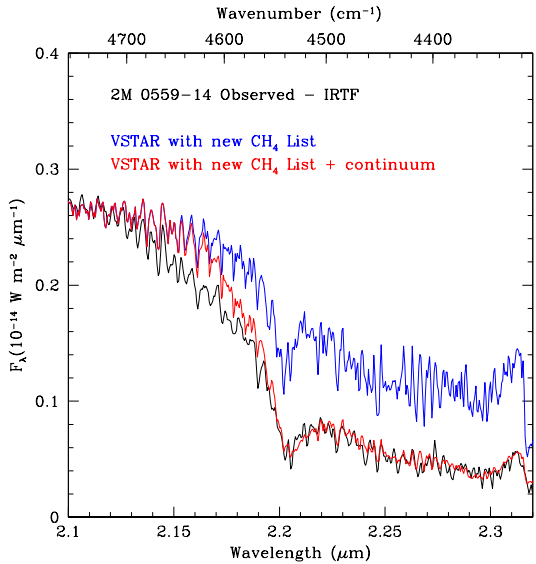}
\caption{The infrared spectrum of 2MASS 0559-14 (T4.5) observed with the IRTF compared to VSTAR model atmospheres containing the new CH$_{4}$ data (effective temperature of 1200 K). Both models are the same, except one model includes the corresponding CH$_{4}$ continuum cross sections.\label{fig5} \vspace{0.2cm}}
\end{figure}

The infrared spectrum of the bright T4.5 dwarf 2MASS 0559-14 has been obtained from the spectral library of the NASA Infrared Telescope Facility \citep{cushing2005, Rayner2009} and has an approximate resolution of 3.8 cm$^{-1}$ ($R \sim 1200$). The spectrum is displayed in Figure~\ref{fig5} alongside model atmospheric spectra. The model spectra are calculated using the Versatile Software for Transfer of Atmospheric Radiation (VSTAR) code and the method outlined in \citet{bailey2012}. The model atmospheres are calculated for an effective temperature of 1200 K and log $g$ = 5 in cgs units \citep{burrows2006}. VSTAR also assumes equilibrium chemistry and recent absorption coefficients for H$_{2}$–-H$_{2}$ collision induced absorption are used \citep{abel2011}. Except for the source of the methane absorption data, the VSTAR model is identical to that described in \citet{yurchenko2014b}. Two model atmospheres have been computed using the new empirical CH$_{4}$ line lists with one also including the CH$_{4}$ continuum cross sections. At model layer temperatures that are between observations, the continuum cross sections are interpolated with a spline. At model layer temperatures that were higher than the experimental observations (i.e, $>$900$^{\circ}$C), the 900$^{\circ}$C continuum cross section was used.

Figure~\ref{fig5} illustrates that our line lists and continua are also able to match low resolution observations and demonstrate the importance of including the continua for brown dwarf comparisons. It is only below about 2.19 $\mu$m that a noticeable difference can be seen. This discrepancy can be explained by the sensitivity of our experimental observations in regions between strong absorption polyad bands and the location of the baseline. The continuum cross sections equal zero for these regions, which are truncated at the noise level. The continuum is a major contributor for low resolution brown dwarf spectra and a minor adjustment for regions below our experimental sensitivity could improve this difference, but further investigations are required to determine the appropriate shape of the continuum between polyad bands. Another factor relates to the 900$^{\circ}$C continuum cross section near 4600 cm$^{-1}$ (2.17 $\mu$m), which can be seen to cross the continua at lower temperatures (Figure~\ref{fig3}). This is inconsistent with absorption cross sections at other temperatures and wavenumbers, where an overall increase is seen. The reason for this overlap is due to a necessary baseline correction. We have tested the effect of increasing the 900$^{\circ}$C continuum for this region and applying this to the VSTAR models; the agreement with observation shows improvement.

A similar match to the same brown dwarf at low-resolution has been obtained using the 10to10 CH$_{4}$ line list \citep{yurchenko2014b}, which contains billions of lines and is therefore time consuming to compute. The models used in this work are very much faster to compute, due to the use of a significantly smaller line list, and because numerous weak lines have been represented as a continuum cross section.

\section{CONCLUSIONS}

The empirical line lists presented in this article are based on the first high-resolution infrared absorption spectra of CH$_{4}$ at high temperatures. Combining the relatively small CH$_{4}$ line lists and quasi-continuum cross sections from this work shows good agreement with high temperature spectra at high and low resolution. This work also demonstrates that theoretical calculations of CH$_{4}$ have significantly progressed in recent years. However, it is the line positions that show the largest disagreements with experiment. It should be noted that these empirical line lists and quasi-continuum cross sections are not intended to replace the comprehensive line parameters from the HITRAN database, but are instead intended to replicate observations at high temperatures. When using these data, the line list and continuum cross section that are closest to observation should be employed. {\color{Black}Users should be aware that our empirical line parameters will be refined as quantum number assignments are made and more sophisticated analyses are carried out in future.}

The scientific potential of high resolution spectra for faint objects, such as exoplanets and brown dwarfs, is only just being realized. For example, cross correlation of high resolution spectra of the star HD 209458 is able to uncover the faint velocity shifted carbon monoxide absorption from the associated exoplanet HD 209458b \citep{snellen2010}. For this method to work with CH$_{4}$, line positions must match laboratory observations precisely and existing calculated spectra are not satisfactory in this regard. When line positions are acceptable for low resolution but not high resolution, this cross correlation method has been shown to fail \citep{hoeijmakers2015}. Over the coming years, new ground and space based telescopes will have greater sensitivity than ever before and it is therefore important to have molecular spectra that can be applied at high resolution as well as low resolution. Currently, only the CH$_{4}$ line lists and continuum cross sections presented here are able to satisfy this requirement.

These line lists and quasi-continua can be used for brown dwarf and exoplanet atmospheric models but they will also help to refine current theoretical calculations. Furthermore, this work demonstrates an alternative way of providing theoretical line lists. It is inevitable that line lists based upon experimental observations will be incomplete due to the sensitivity and resolution limitations, however, the inclusion of a quasi-continuum addresses this problem: effectively accounting for billions of lines beyond detection limits. Representing theoretical line lists in a similar way (i.e., weak and strong components, where the weak lines are used as an absorption cross section) can help to significantly reduce calculation times.

\acknowledgments

Funding for this work was provided by the NASA Planetary Atmospheres Program.

\end{document}